\documentclass[twocolumn,pre,superscriptaddress,aps,nofootinbib]{revtex4}

\usepackage{epsfig}
\usepackage{amsmath}
\usepackage{graphicx}
\usepackage{amsmath}
\usepackage{amsthm}
\usepackage{multirow}
\usepackage{relsize}
\usepackage{amssymb}
\usepackage{bm}
\usepackage{textcomp}

\usepackage{epstopdf}
\usepackage{soul}
\RequirePackage{color}

\DeclareMathAlphabet{\mathitbf}{T1}{cmr}{bx}{it}

\DeclareMathAlphabet{\mathitbf}{T1}{cmr}{bx}{it}

\begin{document}

\title{Equilibrium and out-of-equilibrium critical dynamics of the
  three-dimensional Heisenberg model with random cubic anisotropy}

\author{A. Astillero}
\affiliation{Departamento de Tecnolog\'{\i}a de
  los Computadores y las Comunicaciones, Universidad de Extremadura,
  06800 M\'erida, Spain.}\affiliation{Instituto de Computaci\'{o}n
  Cient\'{\i}fica Avanzada (ICCAEx), Universidad de Extremadura, 06071
  Badajoz, Spain.}

\author{J. J. Ruiz-Lorenzo} \affiliation{ Departamento de F\'{\i}sica,  
  Universidad de Extremadura, 06071 Badajoz, Spain.}
\affiliation{Instituto de Computaci\'{o}n Cient\'{\i}fica Avanzada (ICCAEx),
  Universidad de Extremadura, 06071 Badajoz, Spain.} 

\date{\today}

\begin{abstract}
  We study the critical dynamics of the three-dimensional Heisenberg
  model with random cubic anisotropy in the out-of-equilibrium and
  equilibrium regimes. Analytical approaches based on field theory
  predict that the universality class of this model is that of the
  three-dimensional site-diluted Ising model.  We have been able to
  estimate the dynamic critical exponent by working in the equilibrium
  regime and by computing the integrated autocorrelation times
  obtaining $z=2.50(5)$ (without taking into account scaling
  corrections) and $z=2.29(11)$ (by fixing the scaling corrections to
  that predicted by field theory).  In the out-of-equilibrium regime
  we have focused in the study of the dynamic correlation length which
  has allowed us to compute the dynamic critical exponent obtaining
  $z=2.38(2)$, which is compatible with the equilibrium ones. Finally,
  both estimates are also compatible with the most accurate prediction
  $z=2.35(2)$, from numerical simulations of the 3D site-diluted Ising
  model, in agreement with the predictions based on field theory.
  
\end{abstract}


\maketitle

\section{Introduction}\label{I}

Understanding the influence of disorder on materials is of paramount
importance in both fundamental science and technology.

The effect of one of the most important kind of disorder, the
so-called quenched disorder, where the elements of disorder have
characteristic times significantly longer than the typical
characteristic times of the material's original components, can be
studied using analytical
computations~\cite{goldschmidt:92,holovatch:02,dudka:05b}, numerical
simulations~\cite{ruizlorenzo:22} and
experiments~\cite{cochrane:81,sellmyer:92}.

Furthermore, when focusing on magnetic materials, they can be described,
depending of their anisotropy, by Heisenberg, XY or even Ising
models. On these models, the quenched disorder can be introduced in
different ways, for example, introducing dilution (by means of non
magnetic impurities) in the magnetic material or by inducing uniaxial
anisotropies in the model~\cite{fischer:93}.

In this paper we will focus on the dynamic critical behavior of
the 3D Heisenberg model in presence of (quenched) random cubic anisotropies
(the disorder is localized along the six directions of the cubic
lattice)~\cite{hertz:85,holovatch:02}.

These studies are of special relevance in the understanding of the
physical properties of structurally disordered magnetic
materials. Regarding the technological applications of theses
materials, we can describe the possible use of rare-earth elements as
refrigerants due to the presence of magnetocaloric
effects~\cite{luo:08,luo:09,luo:11}.

From the analytical studies of these models, it is noteworthy that the
perturbative renormalization group (RG) predictions on the critical
behavior of these systems depend strongly on the characteristics of
the disorder.

For the random cubic anisotropy disorder (RAM), RG shows that the
system behaves at criticality like the 3D site-diluted Ising model
(RSIM)~\cite{calabrese:04,dudka:05} with the scaling dimension
(length) of the leading irrelevant scaling field is given by
$\omega_\mathrm{RAM}=-\alpha_\mathrm{RSIM}/\nu_\mathrm{RSIM}$,
being $\nu_\mathrm{RSIM}$ and $\alpha_\mathrm{RSIM}$ the thermal
and specific heat critical exponent of the RSIM~\cite{calabrese:04}.
 
Surprisingly, when the disorder is isotropic, the RG finds no fixed
points: hence, a first order should appears, or there may be no phase
transition at all (see, for example, the discussion of
Ref. \cite{dudka:05b}).
 
In Ref.~\cite{ruizlorenzo:22} the (static) equilibrium critical
phenomena of both isotropic and cubic quenched disorders was
characterized by means of numerical simulations. For the cubic
anisotropy disorder, a second order phase transition in the
ferromagnetic channel raised, but with critical exponents ($\eta$ and
$\nu$) not near those of the RSIM but without the numerical
characterization of the scaling corrections. Furthermore, in the
isotropic disorder, a spin glass phase transition was found. However,
the characterization of the dynamics of these models with both types
of disorder is still lacking.

Given the distinct physics of cubic and isotropic disorders, we focus
in this paper on the characterization of the dynamics at equilibrium
and out-of-equilibrium for the cubic anisotropy disorder.

In order to study the equilibrium critical dynamics, the system
is brought to equilibrium, and subsequently, the integrated
autocorrelation time is calculated.  The scaling of the
integrated autocorrelation times with the lattice size (see below) allow us to
compute the dynamic critical exponent.  The constraint to be in
equilibrium restricts us to simulate relatively small lattice sizes.

In the out-of-equilibrium simulations, we simulate large lattice sizes
and random initial spin configurations. We then track the evolution of
the spin-spin correlation function over the time. Our results
consistently exhibit scaling behavior (see below), enabling us to
characterize the dynamics in this regime. To remain in the
out-of-equilibrium regime for large times it is compulsory to simulate large
lattice sizes.

Both approaches, equilibrium and out-of-equilibrium, should provide
the same dynamic critical exponent which characterizes the critical
dynamics.  We followed this dual approach in the computation of the
dynamic critical exponent of the pure (non-disordered) 3D Heisenberg
model ~\cite{astillero:19}.

The analytical approach to the local relaxational dynamic implemented in the
numerical simulations performed in this paper is the so-called
Model-A dynamics~\cite{tauber:14}. The Model-A dynamics of the
Heisenberg model with cubic anisotropy was study in
Ref.~\cite{dudka:05} using field-theoretical techniques. The outcome
of this analysis is that the dynamic critical exponent is the same as
in the RSIM.

The dynamic critical exponent of the RSIM was
computed analytically, for example, in
Refs.~\cite{blavatska:05,prudnikov:06} and numerically in
Refs.~\cite{parisi:99,hasenbusch:07b}. For future use, we report the
most accurate value: $z=2.35(2)$~\cite{hasenbusch:07b}.

The main aim of this paper is to compute the dynamic critical
exponent, in equilibrium and out-of-equilibrium, in order to check the
theoretical prediction. Therefore, this investigation will provide with the last
critical exponent to fully characterize the universality class of the
3D Heisenberg model with random cubic anisotropy, where the static
exponents $\nu$ and $\eta$ were already computed in numerical
simulations~\cite{ruizlorenzo:22}.

The structure of the paper is the following. In Sec. \ref{MO} we
describe the model and the observables used in the equilibrium and
out-of-equilibrium studies. Next, in Sec. \ref{sec:num}, we report
details on our estimate of the dynamic critical exponent, first in
equilibrium, using the integrated autocorrelation times, and then
out-of-equilibrium by computing and analyzing the dynamic correlation
length. In the last section we will discuss the results and report our
conclusions. Finally, in an appendix, we describe some technical
details on the procedure we employed to calculate some observables in
the out-of-equilibrium dynamics.

\section{The model and Observables}\label{MO}
The Hamiltonian of the three-dimensional Heisenberg model with random cubic
anisotropy is
\begin{equation}
{\mathcal H} =  - J \sum_{<{\mathitbf x},{\mathitbf y}>}  \mathitbf{s}_{\mathitbf{x}}\cdot \mathitbf{s_{\mathitbf y}} -D\sum_{\mathitbf x} (\hat{\mathitbf{n}}_{\mathitbf x}
\cdot\mathitbf{s}_{\mathitbf x})^2 \,.
\label{origham}
\end{equation}
Here, $\mathitbf{s}_{\mathitbf x}$ is a classical three-dimensional
unit vector on the site $\mathitbf{x}$ of a three-dimensional cubic
lattice of size $L$ and volume $V=L^3$ with periodic boundary
conditions, $D>0$ is the strength of the anisotropy, $\hat{\mathitbf
  n}_{\mathitbf x}$ is a (quenched) random unit vector pointing in the
direction of the local anisotropy axis.  The interaction $J>0$ is
assumed to be ferromagnetic, without loss of generality we will fix
$J=1$ hereafter. Finally, the sum $<{\mathitbf x},{\mathitbf y}>$ is
over the six nearest neighbors sites $\mathitbf{y}$ of the site
$\mathitbf{x}$.  The strength of the disorder is controlled by the
ratio $D/J$.

In the cubic anisotropy disorder the vectors $\hat{\mathitbf
  n}_{\mathitbf r}$ point along the six semi-axes of the cubic
lattice with the same probability $1/6$:
\begin{equation}
       p(\hat{\mathitbf{n}})=\frac{1}{6} \sum_{i=1}^3 \left( \delta
    (\hat{\mathitbf{n}}-\hat{\mathitbf{k}}_i)
    + \delta
    (\hat{\mathitbf{n}}+\hat{\mathitbf{k}}_i)\right)\,,
\end{equation}    
where the $\hat{\mathitbf{k}}_i$ ($i=1,2,3$) are the three unit
vectors pointing to the three principal directions of a cubic lattice
and $\delta(\hat{\mathitbf{x}})\equiv
\delta_{\hat{\mathitbf{x}},{\mathrm 0}}$ is the Kronecker delta.

\subsection{Equilibrium}
\label{sec:defequil}

The dynamic critical exponent, within the equilibrium
regime, is computed by analyzing the integrated autocorrelation time in
relation to the lattice size.

For a given observable, we compute ${\cal O}(t)$,  the autocorrelation
function~\cite{madras:89,sokal:92,amit:05} defined as
\begin{equation}
  \label{eq:auto}
  C_{\cal O}(t)\equiv \langle {\cal O}(s) {\cal O}(t+s) \rangle -\langle {\cal O}(t) \rangle^2\,,
\end{equation}
and its normalized version
\begin{equation}
  \label{eq:autoNorm}
  \rho_{\cal O}(t)\equiv \frac{\overline{C_{\cal O}(t)}}
      {\overline{C_{\cal O}(0)}}\,,
\end{equation}
where $\langle(\cdots) \rangle$ is the thermal average
and $\overline{(\cdots)}$ is the average over the disorder.

To overcome the associated bias of a finite number of measurements in
a given sample (notice that we have simulated a large number of MC
sweeps and the associated measurements, see Tables \ref{table:tauD3}
and \ref{table:tauD4}), we have implemented the procedure described in
Ref.~\cite{hasenbusch:07b}.

The integrated autocorrelation time is defined as
\begin{equation}
  \label{eq:tau}
\tau_{\mathrm{int},{\cal O}}= \frac{1}{2}+ \sum_{t=0}^\infty \rho_{\cal O}(t)\,.
\end{equation}

In a data collection with $N$ measurements, the number of
statistically independent measurements of the observable ${\cal O}$ is
just $N/(2\tau_{\mathrm{int},{\cal O}})$~\cite{madras:89,sokal:92,amit:05}.
If the number of
measurements is limited, at large time intervals $t$, the noise will
become more prominent than the signal in $\rho_{\cal O}(t)$. To
address this issue, we employ the following self-consistent approach
to calculate the integrated time
\begin{equation}
  \label{eq:tauM}
\tau_{\mathrm{int},{\cal O}}= \frac{1}{2}+ \sum_{t=0}^{ c \tau_{\mathrm{int},{\cal O}}} \rho_{\cal O}(t)\,,
\end{equation}
where $c$ is usually taken to be 6 or
bigger\cite{madras:89,sokal:92,amit:05}. In this work we have used
$c=10$.

At the critical point, the integrated autocorrelation time of a
long--distance observable exhibits divergence as the system size
increases~\cite{hasenbusch:07b}
\begin{equation}
  \label{eq:tauscaling}
\tau_{\mathrm{int},{\cal O}}\sim L^z \large(1+  O(L^{-\omega})\large)\,.
\end{equation}
Here, $z$ represents the dynamic critical exponent and $\omega$
denotes the leading correction-to-scaling exponent, which corresponds
to the leading irrelevant eigenvalue of the theory.

In this paper, we will focus on the slowest mode offered by the non-local operator ${\cal O}=\mathitbf{M}^2$, where the magnetization ${\mathitbf{M}}$ is defined as
\begin{equation}
{\mathitbf M} = \sum_{\mathitbf x} {\mathitbf s}_{\mathitbf x}\,,
\end {equation}
and we will extract the $z$ exponent by studying its integrated
autocorrelation time.

\subsection{Out of equilibrium}

In the out-of-equilibrium regime, one of the primary observables of
interest is the spin-spin correlation function, defined as
\begin{equation}
  \label{eq:cor}
  C(r,t)=\frac{1}{V} \sum_{\mathitbf x}
  \overline{\langle {\mathitbf s}_{\mathitbf x} {\mathitbf s}_{\mathitbf{r+x}} \rangle_t } \,,
\end{equation}
where $\langle (\cdots) \rangle_t$ stands for the average over
different thermal histories and, as in equilibrium, the horizontal bar
denotes the average over the disorder.

This correlation function satisfies, at criticality, the following
scaling law~\cite{tauber:14}
\begin{equation}
C(r,t)=\frac{1}{r^a} f\left(\frac{r}{\xi(t)}\right) \,,
\end{equation}
which defines the dynamic correlation length, denoted as $\xi(t)$. As
we approach the equilibrium regime, $\xi(t)$ converges to its
equilibrium value.

In three dimensions, at the critical point and in equilibrium, one
should expect
\begin{equation}
C(r,t)\sim\frac{1}{r^{d-2+\eta}} = \frac{1}{r^{1+\eta}} \,,
\label{eq:corre-equi}
\end{equation}
$\eta$ being the anomalous dimension of the field.
  
The correlation length $\xi(t)$ can be estimated by calculating~\cite{janus:08b,janus:09b}
\begin{equation}
 I_k(t)=\int_{0}^{L/2} dr~ r^k C(r,t)\,,
\label{eq:I}
\end{equation}
by means of
\begin{equation}
  \xi_{k,k+1}(t)\equiv \frac{I_{k+1}(t)}{I_k(t)}\,.
\end{equation}

In this work, our focus lies on $\xi_{2,3}$. In spin glasses,
$\xi_{1,2}$ was measured using a correlation function that decays as
$1/r^{0.5}$~\cite{janus:08b,janus:09b}. However, in our case, to
reduce the influence of shorter distances, we have computed higher
values of $I_k$. We refer the reader to
Refs. ~\cite{janus:08b,janus:09b,astillero:19,yllanes:11,young:15} for
a detailed description of the procedure we employed to calculate the
integrals and estimate the statistical error associated with
$\xi_{k,k+1}(t)$ (see Appendix \ref{ApenA} for a detailed description).

Finally, the relationship between the dynamic correlation
length and time is as follows
    
\begin{equation}
  \xi_{k,k+1}(t)\sim t^{1/z} \left(1+ A_k t^{-\omega/z}  \right)\,.
  \label{eq:xiNL}
\end{equation}

\section{Numerical results}
\label{sec:num}

In this section we firstly report the computation of the integrated
correlation time at equilibrium.  After this analysis, we describe our
results in the out of equilibrium regime.

The Metropolis algorithm was used to simulate the model's behavior in
both equilibrium and non-equilibrium simulations. In addition, we have
performed all the numerical simulations using the inverse critical
temperatures provided by Ref.~\cite{ruizlorenzo:22}: $\beta_c=0.7083$
for $D=4$ and $\beta_c=0.6994$ for $D=3$. The starting initial
configuration for the out-of-equilibrium runs was a random
configuration of spins.

To perform the equilibrium numerical simulations we have used two
different algorithms. In the first algorithm (denoted Algorithm A) the
Metropolis algorithm is implemented by modifying the original spin by
adding a random vector (chosen with uniform probability on the surface
of the sphere of radius $\delta$) and normalizing the final vector to
the unit sphere, the magnitude of the random vector $\delta$ is
adjusted in such a way that we maintain an acceptance between 40\% and
60\% (see, for examble, Ref.~\cite{amit:05}).

We have also used a second algorithm (termed as algorithm B) in which
the Metropolis algorithm is carried out by proposing a random spin in
the unit sphere (chosen with uniform probability). Both versions of
the Metropolis algorithm are supposed to belong to the same dynamic
universality class~\cite{sokal:92}.

The equilibrium simulations were executed on CPUs using algorithms A
and B. The out-of-equilibrium simulations were run in CPUs and GPUs
using algorithm A. Details on the implementation of the code in GPUs
can be found in Ref.~\cite{astillero:19}.

\subsection{Equilibrium}
\label{sec:equi}

We have analyzed the integrated correlation times using the
methodology described in the previous section. Notice, that before
performing the measures we need to bring the different systems to the
equilibrium. We have used in this section both versions of the
Metropolis algorithm (algorithms A and B).

In Tables \ref{table:tauD3} and \ref{table:tauD4} we report the values
of the integrated autocorrelation times computed for the different
lattice sizes at the critical point for the two values of the
disorder, $D=3$ and 4, and for algorithms A and B. In addition we show
the number of simulated samples (disorder realizations) and the number
of sweeps (in units of the integrated autocorrelation time) performed
at equilibrium in a given sample.

Notice that, for the same lattice size, the integrated autocorrelation
times are bigger for algorithm A than for algorithm B.  Moreover, the
integrated autocorrelation times are bigger for $D=4$ than for $D=3$,
reflecting that the strength of the disorder induces a slower dynamic.

To compute the statistical error of the integrated correlation time we
have used the jackknife method over the samples~\cite{young:15}.

\begin{table}[h!]
\begin{center}
  \begin{tabular}{c| c c c | c c c}
    \hline \hline
 &\multicolumn{3}{c}{Algorithm A} & \multicolumn{3}{c}{Algorithm B} \\
    $L$ & $\tau_{\mathrm{int},\mathitbf{M}^2}$  &  $n_\mathrm{sw}/ \tau_{\mathrm{int},\mathitbf{M}^2}$ &$n_\mathrm{s}$ & 
        $\tau_{\mathrm{int},\mathitbf{M}^2}$  & $n_\mathrm{sw}/ \tau_{\mathrm{int},\mathitbf{M}^2}$ & $n_\mathrm{s}$\\
  \hline   
8   & 147(1) & 34013 &500 & 44.7(2)  & 111856 &500\\
12  & 368(3) & 13586 & 500 & 105.3(5) & 47483 & 500\\
16  & 670 (6) & 7462 & 500 & 197(1)   & 85380 &  500\\ 
24  & 1723(17) & 5803 & 400 & 490(4) & 20408 & 400\\
32  &  3340(41) & 11976 & 800 &  924(10) & 21640 &  400\\
48  &     -      &   - & -  & 2310(35) & 12987 & 400\\
\hline   \hline
  \end{tabular}
\end{center}
  \caption{ $D=3$. Integrated correlation time of $\mathitbf{M}^2$,
    $\tau_{\mathrm{int},\mathitbf{M}^2}$ for $c=10$, as a function of
    the lattice size, $L$, for the two simulated algorithms. We have
    also reported the number of simulated samples, $n_\mathrm{s}$. We
    also report the length of the run at equilibrium, $n_\mathrm{sw}$,
    in units of $\tau_{\mathrm{int},\mathitbf{M}^2}$.}
  \label{table:tauD3}
\end{table}

\begin{table}[h!]
\begin{center}
  \begin{tabular}{c| c c c| c c c}
    \hline \hline
 &\multicolumn{3}{c}{Algorithm A} & \multicolumn{3}{c}{Algorithm B} \\
    $L$ & $\tau_{\mathrm{int},\mathitbf{M}^2}$  & $n_\mathrm{sw}/ \tau_{\mathrm{int},\mathitbf{M}^2}$ & $n_\mathrm{s}$ & 
        $\tau_{\mathrm{int},\mathitbf{M}^2}$  &  $n_\mathrm{sw}/ \tau_{\mathrm{int},\mathitbf{M}^2}$ & $n_\mathrm{s}$\\
  \hline   
8  &  311 (2) & 16077 & 500 &  59.1(4) & 84602 &500\\
12 &  836(8)  & 5980 & 500 & 148(2)   & 33783 & 500\\
16 &  1685(24) & 2967 &500 & 288(3)  & 17361 & 500 \\
24 &   4547(70) & 2199 & 500  & 765(12) & 13072 & 400 \\
32 &  9523(130) & 4200 & 864 & 1579(21) & 25332 & 692\\
\hline   \hline
  \end{tabular}
\end{center}
   \caption{$D=4$. Integrated correlation time of $\mathitbf{M}^2$,
     $\tau_{\mathrm{int},\mathitbf{M}^2}$ for $c=10$, as a function of
     the lattice size, $L$, for the two simulated algorithms. We have
     also reported the number of simulated samples, $n_\mathrm{s}$. We
     also report the length of the run at equilibrium,
     $n_\mathrm{sw}$, in units of
     $\tau_{\mathrm{int},\mathitbf{M}^2}$.}
  \label{table:tauD4}
\end{table}

\begin{table}[h!]
\begin{ruledtabular}
\begin{tabular}{cccc}
$D$ & Algorithm & $L_\mathrm{min}$ & $z$  \\
  \hline
3 & A & 8 & 2.247(8) \\
3 & B & 12 & 2.22(1) \\ \hline
4& A&  8 & 2.46(1)\\
4 & B & 16 & 2.45(2) \\   
\end{tabular}
\end{ruledtabular}
\caption{Values of $z$ for the different fits assuming no
  correction-to-scaling terms.  For all the fits reported in this
  paper, we have computed the minimum value of $L$ (denoted as
  $L_\mathrm{min}$) , so that, the $p$-value of the least-squares fit
  is greater than 5\%.
  \label{table:resZ}}
\end{table}  

We have analyzed the numerical data reported in Tables
\ref{table:tauD3} and \ref{table:tauD4} neglecting the scaling
corrections (e.g., using $\tau \propto L^z$). The results for the
dynamic critical exponent are reported in Table \ref{table:resZ} for
the four simulated cases. Note that these $z$ estimates are near the
RSIM dynamic exponent $z=2.35(2)$, but a bit smaller
for $D=3$ and a bit larger for $D=4$. This results also 
support that both algorithms belong to the same dynamic universality
class.

Moreover, assuming universality, the four cases should present the
same critical behavior and to show this fact, it is compulsory to take
into account the corrections to scaling (the same $z$ exponent with
the same scaling corrections)
~\cite{ballesteros:98b,hasenbusch:07,hasenbusch:07b,hasenbusch:08b,fernandez:16b,fytas:13,fytas:16,fytas:19}.

To check this scenario, we have tried a joint fit of the four cases
using the same $z$ exponent and the same field-theory result for the
leading irrelevant scaling field
$\omega_\mathrm{RAM}=-\alpha_\mathrm{RSIM}/\nu_\mathrm{RSIM}=
3-2/\nu_\mathrm{RSIM}=0.072(6)$~\cite{calabrese:04} where we have used
$\nu_\mathrm{RSIM}=0.683(2)$~\cite{hasenbusch:07}: this is a very small
value for the $\omega$ exponent, so we are confronting strong scaling
corrections.

From this fit, we obtained $z=2.29(11)$ with 
$\chi^2/N_\mathrm{DOF}=18/10$ and a associated $p$-value=0.055.\footnote{The
  goodness of the fit is reported via $\chi^2/N_\mathrm{DOF}$,
  $N_\mathrm{DOF}$ stands for the number of degrees of freedom of the
  fit, and the associated $p$-value is also denoted as
  $Q$~\cite{young:15}.} To have an acceptable $p$-value we have used
  the $L\ge 8$ data for the Algorithm A and the $L\ge 12$ data for the
  algorithm B.

Taking into account the uncertainty in the value of $\omega$ 
provides an additional uncertainty of $5\times 10^{-3}$ in the value of $z$, which is neglibigle in confronting with the computed error (0.11).
Hence, our final estimate is: $z=2.29(11)$.

\begin{figure}[h!]
\centering
\includegraphics[width=\columnwidth, angle=0]{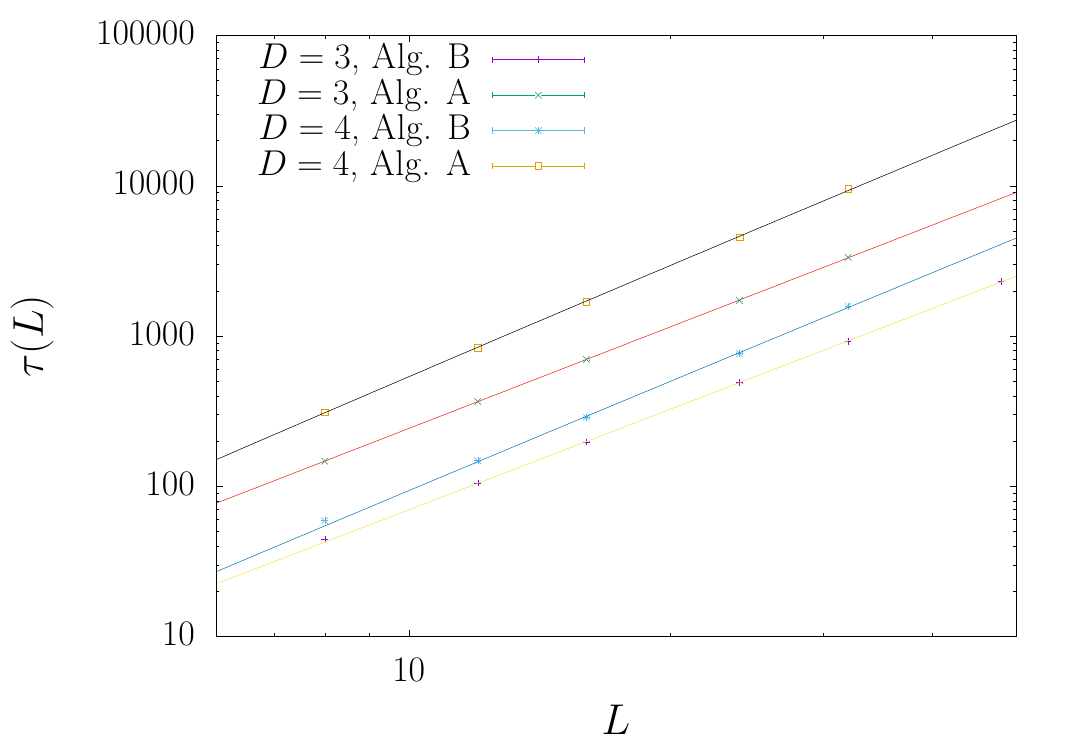}
\caption{(color online) Behavior of the integrated correlation time
  $\tau_{\mathrm{int},\mathitbf{M}^2}$, as a function of the lattice
  size, $L$, for the two simulated algorithms (A and B) and for $D=3$
  and $D=4$. We have also shown our best joint fit taking into account
  corrections to the scaling (see the text).}
\label{fig:equil}
\end{figure}

\subsection{Out-of-equilibrium: Correlation length}
\label{sec:zeta}

In this section we compute the dynamic critical exponent using the
correlation function computed in an out-of--equilibrium simulation.

\begin{figure}[h!]
\centering
\includegraphics[width=\columnwidth, angle=0]{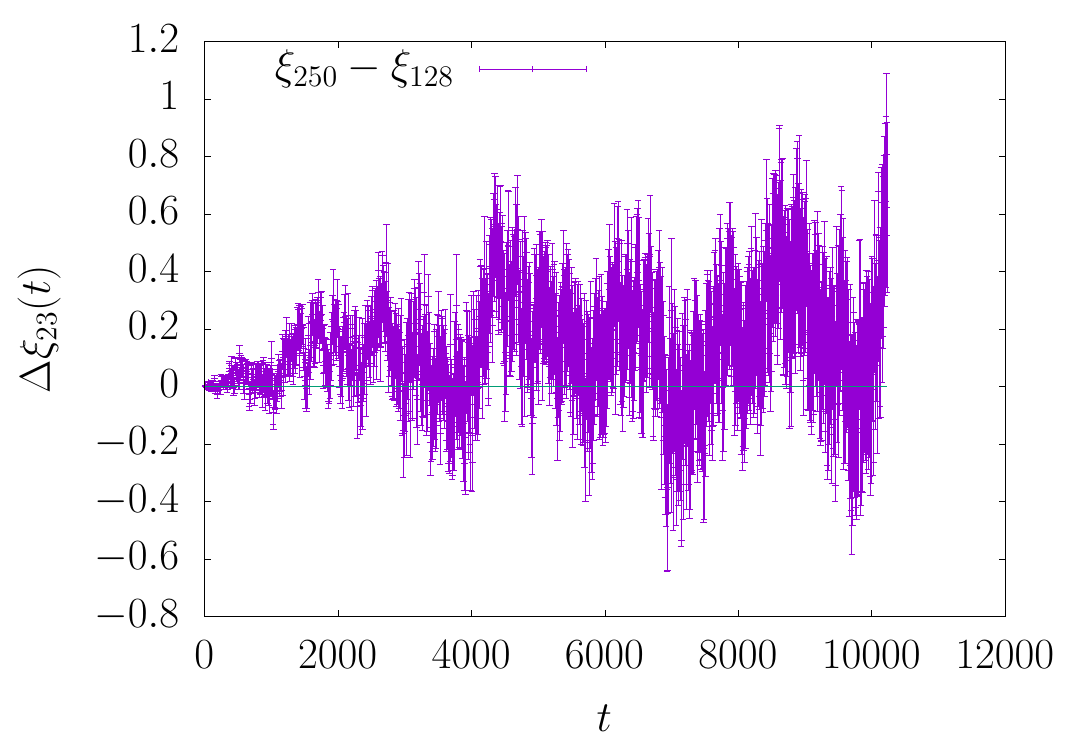}
\caption{(color online) Behavior of $\xi_{23}(t,L=250)-\xi_{23}(t,L=128)$
  versus time for $t$ and for $D=3$. The horizontal green line marks the zero.}
\label{fig:offequilCOMP}
\end{figure}

Firstly, we have studied the finite size effects of the
out-of-equilibrium numerical simulations. To do that, we have computed
the dynamic critical correlation length, $\xi_{23}$, using $L=128$
and $L=250$ sizes. In Fig. \ref{fig:offequilCOMP} we show the
difference between the correlation length ($\xi_{23}$) computed
simulating both lattice sizes, and we can assess that (in the
statistical errors) both estimates of $\xi_{23}$
are compatibles as far as the Monte Carlo time is less
than $10^4$ Monte Carlo sweeps. Hence, our out-of-equilibrium analysis
will be based on the $L=128$ lattice for which we have simulated a
larger number of samples. We report in Table \ref{table:zeta} the
number of samples we have simulated in the $L=128$ lattice for the two
values of the cubic anisotropy.\footnote{In the $L=250$ simulations we
have simulated 1000 samples for $D=3$ and $1400$ for $D=4$.}

\begin{figure}[h!]
\centering
\includegraphics[width=\columnwidth, angle=0]{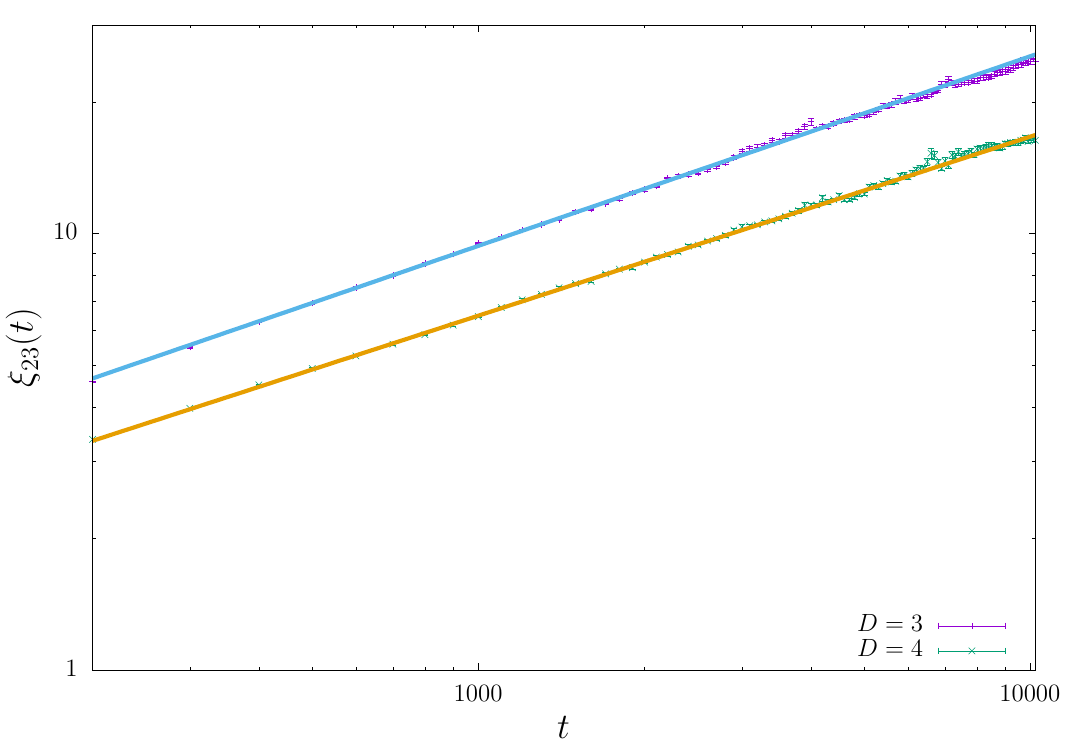}
\caption{(color online) Behavior of $\xi_{23}$ versus time for $L=128$
  and for $D=3$ and $D=4$. The straight lines are the fit, in the
  interval $[1000,6000]$ for the pure law behavior, see Eq.
  (\ref{eq:xiNL}), the described in the text.}
\label{fig:offequil}
\end{figure}

Figure \ref{fig:offequil} shows the behavior of the dynamic
correlation length ($\xi_{23}$) as a function of time for a $L=128$
lattice and $D=3$ and $D=4$. The slope of the curves (in a double
logarithmic scale) provided us with the estimate of $1/z$, neglecting
scaling corrections, and we report our estimates for $z$ in Table
\ref{table:zeta}. As in the equilibrium analysis, we have used the
jackknife method over the samples to compute the statistical errors of
the different observables, including the dynamic critical
exponent~\cite{astillero:19}.

Our final estimate is $z=2.35(3)$ for $D=3$ and a slightly higher
value (but statistically compatible) $z=2.44(4)$ for $D=4$. The
weighted average of these two values is $z=2.38(2)$.

In Appendix \ref{ApenA} we report the detailed procedure we have
followed in order to compute the statistical errors on the dynamic
correlation length and on the $z$ exponent.

\begin{table} [h!]
\begin{ruledtabular}
\begin{tabular}{cccc}
$D$ &  $L$ & $z$ & $n_\mathrm{s}$  \\
  \hline
3  & 128 & 2.35(3) &  4000\\ 
4  & 128 & 2.44(4) &  1600\\ 
\end{tabular}
\end{ruledtabular}
\caption{Dynamic critical exponent from the scaling of the dynamic
  correlation length for $D=3$ and $D=4$ using $\xi_{23}(t)$.
  \label{table:zeta}}
\end{table}  

\section{Conclusions}
\label{sec:concl}

By computing the dynamic critical exponent $z$ we have fully
characterized the universality class of the three-dimensional
Heisenberg model with cubic anisotropy by complementing the static
critical exponents computed in Ref. \cite{ruizlorenzo:22}.

We have computed the dynamic critical exponent by using equilibrium
and out-of-equilibrium approaches.

In the equilibrium simulations, we have analyzed the behavior of the
integrated correlation times as a function of the lattice size. In the
out-of-equilibrium simulations, we have computed the dynamic
correlation length and we have studied its behavior as a function of
the Monte Carlo time.

In order to asses universality we have analyzed, at equilibrium, the
system for two values of the anisotropy coefficient $D$, namely, $D=3$
and 4, and by using two different implementations of the Metropolis
algorithm.

We have obtained $z=2.29(11)$ for the equilibrium analysis doing a
joint fit of the numerical behavior of the integrated correlation time
for both values of $D$ and for both algorithms and taking into account
the leading scaling correction provided by the field theoretical
analysis of the model~\cite{calabrese:04} ($\omega=0.072(6)$).  The
comparison of the computed value with that of the RSIM is good:
$z=2.29(11)$ versus $2.35(2)$.

Notice we have a large statistical error (a factor 6) in confronting
with that of the RSIM. It is important to remark that in the
equilibrium numerical studies of the Random Anisotropy model with
cubic anisotropy we have simulated vector spins (instead of Ising
ones) and we have not at our disposal clusters methods (to generate
the equilibrated initial conditions, which will be evolved using a
local Monte Carlo algorithm, the Metropolis algorithm in our
case). Finally, only assuming the theoretical prediction, we can fix
the scaling correction in the analysis of the equilibrium
autocorrelation times.

The analysis of the out-of-equilibrium data is much more involved
(mainly due to the statistical correlations of the data at different
Monte Carlo times)~\cite{astillero:19}, and we have not been able to take into
account the scaling corrections in the fits and to perform a joint fit
for both values of the anisotropy coefficient $D$ (to check
universality in the out-of-equilibrium regime). Despite these two
issues, we have obtained values fully compatible with the equilibrium
one, being our final value for the out-of-equilibrium analysis
$z=2.38(2)$.

The field-theoretical approach of the renormalization group shows that
the universality class of the 3D Heisenberg model with cubic
anisotropy is that of the RSIM for both static and
dynamic. Equilibrium numerical simulations reported static critical
exponents not so compatible with those of the RSIM, $\eta=0.03(1)$ and
$\nu=0.737(8)$~\cite{ruizlorenzo:22} to be confronted with those of
the RSIM: $\eta= 0.036(1)$ and $\nu=0.683(2)$\cite{hasenbusch:07}. We
remark that in Ref. ~\cite{ruizlorenzo:22}
was not possible to characterize the leading scaling
corrections, and that only the exponents for the crossing point of the
two larger lattice were reported.

In summary, we have presented numerical evidences that the equilibrium
dynamical behavior fit pretty well with the analytical predictions: i)
our estimate of the
dynamic critical exponent is compatible with that of the diluted
ising model ($z=2.35(2)$~\cite{hasenbusch:07b}) and ii) the scaling
corrections could be parametrized using the results of
Ref. \cite{calabrese:04}. The analysis of our out-of-equilibrium
numerical simulations provides with a dynamic critical  exponent compatible with
that of the diluted Ising model, but, we have been unable to control
the scaling corrections.

An open problem is to determine, by simulating larger lattices,
whether the static exponents (for example, those obtained using the
quotient method) can be described with $\omega=0.072$~\cite{calabrese:04}, and whether their
extrapolation to infinite volume is compatible with the RSIM static
exponents.

\section{Acknowledgments}

We thank M. Dudka, A. Maiorano, V. Martin-Mayor and Y. Holovatch for
discussions.

We also thank an anonymous referee for suggestions and comments.

We acknowledge financial support from Grant No. PID2020-112936GB-I00
funded by MCIN/AEI/10.13039/501100011033, and from Grant No. GR24022
funded by Junta de Extremadura (Spain) and by European Regional
Development Fund (ERDF) “A way of making Europe.”

We have performed the numerical simulations in the computing
facilities of the Instituto de Computaci\'{o}n Cient\'{\i}fica
Avanzada (ICCAEx) and in the Extremadura Research Centre for Advanced
Technologies (CETA-CIEMAT), funded by the European Regional
Development Fund (ERDF). CETA-CIEMAT belongs to CIEMAT and the
Government of Spain.

\appendix

\section{Computation of the dynamic correlation length and the  $z$ exponent with their associated statistical error.}
\label{ApenA}

We outline the procedure followed to compute $\xi(t)$ and its
associated dynamic critical exponent
$z$\cite{lulli:15,janus:08b,janus:09b,astillero:19,yllanes:11,young:15,michael:94}. A
key aspect of our approach is to avoid the use of the full covariance
matrix, as it is often singular (see,
e.g.,\cite{yllanes:11,seibert:94}). The method proceeds as follows:

\begin{enumerate}

\item The statistical error of $C(r,t)$ is estimated using the
  jackknife method over the set of initial conditions.

\item To compute $I_k$, we introduce a cutoff to ensure reliable
  signal-to-noise ratios in $C(r,t)$ at large distances:

    \begin{itemize}

    \item The cutoff $\Lambda$ is determined by the condition $\sigma[C(\Lambda,t)] = 4,C(\Lambda,t)$.

\item For a fixed time $t$, and over the range $r_\mathrm{min} < r <
  \Lambda$, the correlation function is fitted to the
  form:
  \begin{equation}
    C(r,t) = \frac{a_1}{r^{a_2}} \exp(-a_3
    r^{a_4})\,,
    \label{eq:corre}
  \end{equation} where $r_\mathrm{min}$
  is the minimum distance for which a good fit to Eq.\eqref{eq:corre}
  is achieved, typically satisfying that the diagonal
  $\chi^2/\mathrm{d.o.f.} \sim 1$.

\item In order to compute $\xi(t)$, the integral in Eq.\eqref{eq:I} is evaluated numerically using
  the measured values of $C(r,t)$ for $r < \Lambda$, and the fitted
  values from Eq.\eqref{eq:corre} for $\Lambda < r < L/2$.

\item The statistical error in $\xi(t)$ is again computed via the
  jackknife method over the initial conditions. The time interval used
  in the fit is chosen by requiring that the diagonal
  $\chi^2/\mathrm{d.o.f.} \sim 1$.

    \end{itemize}

  \item To compute the dynamic critical exponent, we perform the
    following two steps:
    
    \begin{itemize}
\item The values of $\xi(t)$ are then used to determine $z$ using the
  diagonal covariance matrix of the correlation length based on the
  jackknife statistical errors of the correlation length.

\item The associated error of $z$ is estimated from the standard
  deviation across jackknife blocks of the correlation length. The
  extraction of $z$ within each (jackkife) block employs the diagonal
  covariance matrix. Despite this simplification, the jackknife
  approach accurately captures correlations between different time
  points.

  It is important to note that the diagonal
     $\chi^2/\mathrm{d.o.f.}$ does not have the same statistical
     interpretation as the full (non-diagonal) version. As discussed
     in detail in Sec. B.3.3.1 of Ref.~\cite{yllanes:11}, this metric
     behaves as though there are fewer degrees of freedom, and
     therefore, conventional confidence intervals cannot be reliably
     computed.

     Furthermore, Ref.~\cite{lulli:15} demonstrates that the error bars
     obtained via this jackknife method—ignoring inter-data
     correlations—are essentially equivalent to those computed when
     such correlations are fully considered.
     \end{itemize}
     
\end{enumerate}


\end{document}